\documentclass[12pt,preprint]{aastex}

\newbox\grsign \setbox\grsign=\hbox{$>$}
\newdimen\grdimen \grdimen=\ht\grsign
\newbox\laxbox \newbox\gaxbox
\setbox\gaxbox=\hbox{\raise.5ex\hbox{$>$}\llap
      {\lower.5ex\hbox{$\sim$}}}\ht1=\grdimen\dp1=0pt
\setbox\laxbox=\hbox{\raise.5ex\hbox{$<$}\llap
      {\lower.5ex\hbox{$\sim$}}}\ht2=\grdimen\dp2=0pt
\def\gax{\mathrel{\copy\gaxbox}}

\begin{document}

\title{
A Direct Upper Limit on the Density of Cosmological Dust from the
Absence of an X-ray Scattering Halo around the ${\it z}
= 4.3$ QSO 1508+5714
}

\author{Andreea Petric, Gisela A. Telis, Frits Paerels, and David J.
Helfand
}

\affil
{Columbia Astrophysics Laboratory
and Department of Astronomy, Columbia University,
550 W. 120th St., New York, NY 10027}

\begin{abstract}
We report on the results of a search for an intergalactic X-ray dust scattering halo
in a deep observation of the bright, high-redshift quasar QSO
1508+5714 with the {\it Chandra} X-ray Observatory. We do not
detect such a halo. Our result implies an upper limit on the density of 
diffuse,
large-grained intergalactic dust of $\Omega_{\rm dust} < 2 \times
10^{-6}$, assuming a characteristic grain size of $\sim 1 \mu$m.
The result 
demonstrates the sensitivity of this technique for detecting very small
amounts of intergalactic dust which are very hard to detect otherwise.
This will allow us to put important constraints on systematic effects
induced by extinction on the interpretation of the SN Ia Hubble Diagram,
as well as on the amount and 
properties of cosmological
dust being expelled into the intergalactic medium at early ($z\geq 2$) 
times.
\end{abstract}

\section{Introduction}
	Recent discoveries of large quantities of dust at high 
redshifts, such as that of  $\sim 7\times 10^8$ 
~M$_{\odot}$
of dust in the $z~=~6.42$ redshift QSO J1148+5251 (Bertoldi et al. 2003)
and in other high redshift ($z ~\geq~4$) sources (e.g. Priddey et al. 
2003),
suggest that substantial amounts of dust had been already produced at 
times
as early as those corresponding to $z \sim 6$.
One possible technique for detecting and quantifying the amount of dust
grains produced at early times, and their dispersal into the 
intergalactic
medium, is to search for X-ray scattering halos around bright 
extragalactic X-ray sources.

Dust particles will scatter X-rays, and dust scattering halos are 
indeed
seen routinely around distant Galactic X-ray sources.
The absence of detectable systematic reddening of the SN Ia with
increasing redshift means that any intergalactic dust extinction must 
be grey at optical wavelengths,
and it suggests that if dust does exist in large quantities at these 
redshifts,
the grains must be large, that is at least larger than a wavelength of
red light. This circumstance highly favors
the formation of an X-ray dust scattering halo: the integrated X-ray 
scattering
cross section per particle increases as the fourth power of the 
particle diameter
(Mauche \& Gorenstein 1986, and references therein). Consequently,
sensitive X-ray observations can provide an independent measurement of 
the
dust density, and the observations are sensitive precisely to a population of grains
that would elude optical and near-infrared detection. 
A search for intergalactic scattering halos was first
suggested by Evans, Norwell, \& Bode (1985).

The X-ray scattering is not a small effect. A rough estimate of the
magnitude of the effect may be obtained from the following simple 
argument. The
total scattering cross section (assuming spherical particles of
density $\rho$ and radius $a$) scales approximately as $\sigma_{\rm
scattering} \propto \rho^2 a^4 E^{-2}$, with $E$ the photon energy,
  while the characteristic angular width of the cross section, $\theta$,
is approximately $\theta \approx 10
(E/{\rm 1\ keV})^{-1} (a/0.1\ \mu{\rm m})^{-1}$ arcmin
(Mauche \& Gorenstein 1986).

For small optical depth $\tau_{\rm scattering}$, the fractional halo
intensity $f \equiv I_{\rm halo}/(I_{\rm halo}+I_{\rm unscattered})$
is approximately equal to $f \approx \tau_{\rm scattering}$. We can
calculate the scattering optical depth through the Universe by
integrating
\begin{eqnarray}
d\tau_{\rm scattering} & = & \sigma_{\rm scattering}~n_{\rm dust}~
dl \approx \cr
\medskip
& \approx & 6.5 \times 10^{-2}~h_{75}
\left({{\Omega_d}\over{10^{-5}}}\right) \times \cr
&& \times
\left({{\rho}\over{3~{\rm g~cm^{-3}}}}\right)
\left({{a}\over{1~\mu{\rm m}}}\right)
\left({{E}\over{1~{\rm keV}}}\right)^{-2}
	{{(1+z)^2}\over{E(z)}}~dz,
\end{eqnarray}
with $\Omega_d$ the dust mass density in units of the critical
density, $h_{75}$ the Hubble Constant in units 75 km s$^{-1}$
Mpc$^{-1}$, and $E(z) \equiv \left\{\Omega_m(1+z)^3 + \Omega_R(1+z)^2
+ \Omega_{\Lambda}\right\}^{1/2}$. Here $\Omega_m$, $\Omega_R$, $\Omega 
_{\Lambda}$
  stand for the density of baryonic matter, curvature, and dark energy
  respectively as a fraction of the critical density.
  We have assumed a constant comoving
dust density; a weak dependence of the optical depth on the
composition of the dust has been suppressed.
For $z >> 1$, assuming $\Omega_R = 0$ (de Bernardis et al. 2000), we find
\begin{eqnarray}
\tau_{\rm scattering} & \approx &
4.3 \times
10^{-2}~\Omega_m^{-1/2}~\left((1+z)^{3/2}-1\right)~h_{75}
\left({{\Omega_d}\over{10^{-5}}}\right) \times \cr
&& \times
	\left({{\rho}\over{3~{\rm
	g~cm^{-3}}}}\right)
	\left({{a}\over{1~\mu{\rm m}}}\right)
	\left({{E}\over{1~{\rm
	keV}}}\right)^{-2},
\end{eqnarray}
so for $z \sim 4$, $f \sim 0.44 ~\Omega_m^{-1/2}~
(\Omega_d/10^{-5})~(a/1~\mu{\rm m})
(\rho/3\ {\rm g\ cm^{-3}}) \times
(E/{1~{\rm keV}})^{-2}$. This shows that
a moderately deep image of a sufficiently distant point source can
place significant constraints on the dust density. The optical depth
does not depend very strongly on the precise grain shape, for fixed cosmological dust density and characteristic grain radius. This is
intuitively obvious: needle-shaped grains of radius $a$ can be thought
of, to lowest order, as just strings of spheres of radius $a$. 
It is conceivable that the dust population is dominated by long, 
small-diameter needles, with physical dimensions such that the 
X-ray scattering effect is suppressed, for a given amount of optical extinction.
For simplicity, in the absence of significant optical reddening and a lack of information
on the shapes of grains in the IGM, we will perform our analysis assuming a
spherical geometry for the grains. 

Here, we report on the analysis of a dedicated deep exposure with {\it
Chandra} on the bright, distant ($z = 4.30$) quasar QSO1508+5714
(Moran \& Helfand 1997, and references therein), designed to set
constraints on the density of intergalactic dust.

\section{Data Analysis}

QSO1508+5714 was observed with {\it Chandra} on June 10, 2001,
for 88971 sec
(effective exposure time), with the ACIS/S3 chip at the focus.
The data were processed with {\it Ciao} 2.2.1. We reprocessed the
data with the latest ACIS/S3 gain map, retaining only event grades
0,2,3,4,6. The background was stable throughout our observation, and we
retained the entire exposure for analysis. A close examination of the
image of the source shows that it is definitely extended (Figure 1).
As is apparent from Figure 1, however, the extended feature
contains $< 10 \%$ of the source counts,
and does not
affect the radial profile of the source beyond $\sim 4\arcsec$, so
will not affect our measurement of a scattering halo. This jet,
the highest-redshift X-ray jet known, is discussed elsewhere 
(Siemiginowska et al. 2003).
In order to minimize the
background, we excluded events outside the energy range between 300 and
8000 eV (the instrumental background in the S3 chip rises steeply
outside these boundaries). We used the {\it Ciao} tasks
{\tt celldetect} and
{\tt dmfilth} to find and eliminate point sources in the image; only
5 weak sources were found within 3 arcmin of the quasar position.
We measured the surface brightness in the image
in a series of concentric annular
regions centered on the peak brightness of the image (which coincides
with the optical position of the quasar to within 0.5 arcsec in RA
and declination; Hook et al. 1995).

Our attempt to detect a faint extended scattering halo obviously
requires a careful treatment of the background. We retrieved the
standard ACIS/S3 quiescent background data
({\tt acis7sD2000-12-01bkgrndN0001.fits}, 330 ksec exposure),
and reprocessed
it with the same gain map we used for the quasar image. We restricted
event energies to the range $300-8000$ eV, and re-projected the
background events using the aspect solution provided by the CXC for our
observation. As a test, we subtracted a smoothed version of the
resulting background image, simply scaled by exposure time,
from the quasar image, and found that the
residual intensities were generally of order 1$\% $ of the background
image, or less. A very faint diffuse structure (roughly circular shape,
diameter $\sim$ 5 arcsec) remains, located at position angle 70
degrees (measured East from North) at 25 arcsec
from the quasar, with a peak surface brightness of 0.2 counts 
arcsec$^{-2}$.
The total flux in this structure is less than 0.1$\% $ of  
the quasar flux, so we did not subtract it from the image.
Otherwise, the smoothed background-subtracted image appears completely
dark outside a circle 20
arcsec in radius centered on the quasar.
The background is slightly spatially inhomogeneous
(variations up to 10$\% $ from the mean surface brightness
occur near the position of the quasar, on scales of $\sim$ 1 arcmin),
  so we chose to use the same set of
annular extraction regions as we used for the quasar to extract a
background histogram. This histogram was subtracted from the radial
intensity profile of the quasar. The result is shown in Figure 2. We
have not made corrections for the angular dependence of the
telescope throughput, since these remain small (a few percent) over
the angular range of interest.

The wings of the {\it Chandra} point source response are dominated by
scattering by microroughness on the mirrors. The amplitude and angular
distribution of the mirror-scattered light depend strongly on photon
energy, and we therefore need to
carefully model the wings in the image of the quasar.
In order to investigate the intensity distribution,
we compared the radial profile to the image of a low-redshift
extragalactic point source with a spectrum similar to
QSO1508+5714. We selected the ACIS/S3 image of 3C273 (obsid 1712);
at redshift $z = 0.158$, this source is too nearby to have a measurable
intergalactic dust halo. Its Galactic interstellar column density is
small ($N_{\rm H} \approx 1.7 \times 10^{20} $
atoms cm$^{-2}$), so that it does not
have a Galactic dust scattering halo brighter than a few percent
of the point
source flux, either; moreover, the Galactic scattering halo will be
very extended and have very low surface brightness.
We followed the same general procedures for the
extraction of a radial intensity profile for 3C273 as we used for
QSO1508+5714.
There are no bright point sources near 3C273, and the object
is so bright that we simply determined a constant background from a
region located well away from the image of 3C273. When we calculated
the radial profile for 3C273, we took care to exclude angular sections
centered on the jet (Marshall et al. 2001; Sambruna et al. 2001)
and the two detector readout streaks.

Scaling the profile for 3C273 to that of QSO1508+5714 requires a little
care. 3C273 is so bright that the center of the image is depressed due
to pileup
\footnote{{\it Chandra} Proposers Observatory Guide v5.0, Sect. 6.16}.
We therefore ignored the profile inside 8 arcsec, as being
affected by pileup.
We scaled the profile of 3C273 to that of our
quasar over the angular range $9-100$ arcsec;
the result is displayed in Figure 2. As can be seen, the wings
of the two profiles outside a radius of $\sim 10$ arcsec
coincide closely.

We performed a raytrace simulation with {\tt MARX}
for comparison with the core of the profile of QSO1508+5714. We fit the
spectrum of the quasar with a power law over the range 0.3-8 keV,
finding a photon index $\Gamma
= 1.53$, and column density $N_{\rm H} = 8.4 \times 10^{20}$ cm$^{-2}$.
These parameters were used for input into {\tt MARX}. For comparison
with the 1995 {\it ASCA} observation (Moran \& Helfand 1997), we also
record the results of a power law fit with photon energy restricted to
$E > 0.5$ keV: we find
$\Gamma = 1.44 \pm 0.05$, $N_{\rm H} = (4.8 \pm 1.4) \times 10^{20}$
cm$^{-2}$. The power law slope is identical to the {\it ASCA} value;
the column density is somewhat larger, but this
depends on the accuracy of the {\it ASCA} and
ACIS/S calibrations below 0.8 keV.
The flux in the observed
$0.5-10$ keV range in this model is $(5.1 \pm 0.4) \times 10^{-13}$
erg cm$^{-2}$ s$^{-1}$.
We assumed a simple point source on--axis (the
actual off-axis angle of the source is 38 arcsec, small enough that it
should not distort or broaden the PSF with respect to the on--axis
performance).
The resulting radial intensity profile is overlaid on the data in
Figure 2. Inside 3 arcsec, the quasar profile deviates from that of a
point source, as noted above; we therefore scaled the simulated profile
by eye to the measured profile, matching the amplitudes between
9 and 20 arcsec. At large radii, it appears that the simulation
slightly under-predicts the amplitude of the scattering wings observed
in 3C273, an effect that has been noted during flight calibration
(Gaetz 2002). The agreement between the quasar and the combined
3C273/model profiles is good, and we conclude that there is no
evidence for a substantial extended X-ray scattering halo.

\section{Modeling the halo}
We estimated the amount of scattered emission as a function of 
observed angle
using the angular scattering cross-sections derived in the
Rayleigh--Gans approximation for a spherical particle of density $\rho$
and radius $a$, (e.g. Mauche $\&$ Gorenstein 1986):
\begin{equation}
{{d\sigma}\over{d\Omega}}~=~9.3~\times 
~10^{-14}{\left({2Z}\over{M}\right)}^2{\left({\rho}\over{3\ {\rm g\ cm^{-3}}}\right)}^2 
\left({{a}\over{0.1\mu m}}\right)^6
~{\left[{F(E)}\over{Z}\right]}^2 ~ 
\exp\left(-{\theta_{\rm scat}^2}\over{2\tilde{\sigma}^2}\right)~
{\rm{cm}^{2}\ \rm{arcmin}^{-2}}
\end{equation}
where $F(E)$ is the atomic scattering factor, $Z$ is the atomic 
charge, and $M$ the atomic mass number.
$E$ is the energy of an incoming photon as seen by the grain, so
$E~=~E_0{\left({1~+~z_g}\over{1~+~z_s}\right)}$ with $E_{0}$ the 
photon energy at
the quasar, and $z_s$ and $z_g$ the redshifts of the source and the grain,
respectively. Photons are scattered through an angle $\theta_{\rm scat}$, and the rms scattering angle is $$\tilde{\sigma}~=~{{10.4}\over{E({\rm keV})~a(0.1~\mu 
m)}}~\rm{arcmin}.$$ The monochromatic intensity of the scattered
radiation $I_{\rm{scattering}}$ as a function of angle $\theta$ on the sky is given by
an integral along the line of sight:
\begin{equation}
I_{\rm scattering}(\theta )=\int _0 ^{z_s}
{{L_0}\over{(1+z_s)4\pi~r_{sg}^2}}~
{{d\sigma}\over{d\Omega}}
\exp(-\tau_{\rm{scattering}})~n_g dl
\end{equation}
Here, $n_g$ is the volume number density of grains, $L_0$ is the monochromatic luminosity of the quasar (in erg s$^{-1}$ keV$^{-1}$, in the restframe), and $\tau_{\rm scattering}$ is the scattering optical depth
along the line of sight from observer to quasar. The differential cross section $d\sigma/d\Omega$ is to be evaluated at the redshift of the grain. 
Finally, $r_{sg}$ is the comoving distance between intervening 
grains and the quasar:
\begin{equation}
r_{sg}= {c\over H_0} \int _{z_g} ^{z_s} {{dz}\over{E(z)}}.
\end{equation} 
We evaluate the integral in Eq.(4) numerically. Throughout, we assume $\Omega_m = 0.3, \Omega_{\Lambda} = 0.7, \Omega_R = 0$.

For an assumed grain size distribution, grain density and
composition, and space distribution of the dust (all of
which may evolve with redshift), one can in principle exactly
calculate the intensity and shape of the scattering halo. Fitting such
model halos to the data would yield the most constraining upper limits on
the overall density of the dust. Here, we report an
upper limit that is robust, and displays the dependence on the
overall dust parameters as explicitly as possible, based on a set
of simplifying assumptions.

In the first simplification, we assume that the dust has constant
comoving density.  This choice is motivated in part by
detection of thermal dust emission from high redshift QSOs selected 
from
the SDSS survey out to redshifts of 6, and reobserved at mm wavelengths
(Omont et al. 2001, 2003, Carilli et al. 2001, Bertoldi \& Cox 2002, 
Bertoldi et al. 2003).
Also, Pettini et al. (2003) find that the metallicity of the Lyman 
$\alpha$ forest appears to show little evolution out to redshifts $z \sim 5$. This 
suggests that dust is already present
in significant amounts at redshifts larger than 1.
Secondly, we assumed that the dust is
homogeneously distributed. This is not a severe restriction,
since the Universe is evidently optically thin to dust scattering, so
the total amount of scattered light is not sensitive to the precise
distribution of the scattering medium. We will also be studying a
relatively large area on the sky (of order a few square arcminutes),
and since there should be of order 10 star forming galaxies per
square arcmin out to high redshift as possible dust sources (Adelberger \& Steidel 2000;
Blain et al. 2002), effects of cosmic variance should not be
dominant.

Given the probable blazar-like
nature of the quasar (its isotropic luminosity in the rest frame
1.5--10 keV band is $1.0 \times 10^{47}$ erg s$^{-1}$, for $H_0 = 50$
km s$^{-1}$ Mpc$^{-1}$, $q_0 = 0$),
one might worry that the radiation is strongly beamed in our
direction, and that no photons are emitted at sufficiently large angles
from the line of sight to be scattered into our direction. However,
the bulk
Lorentz factors in X-ray jets are believed to be of order $5-10$,
which means that typical opening angles are of order several degrees,
much larger than an X-ray halo; thus, beaming is unlikely to suppress
the halo significantly.

Photon travel time delays could also in principle affect the shape and
intensity of the halo. But typical time delays will be of order
$\theta_{\rm scattering}^2$ times the light travel time to the quasar,
so for scattering angles of order an arcminute, all that is required
is that the quasar X-ray emission has been stable on the average
for the last few thousand years, which is much shorter than the
estimated average length of the accretion phase in the life of a
quasar. The halo intensity
depends on the past luminosity
history of the quasar, and so the fractional halo
intensity we measure with
respect to the flux of the quasar does fluctuate
with the instantaneous
quasar flux. Between the {\it ASCA} observation in
1995 (Moran and Helfand 1997), and the present observation with {\it
Chandra}, the object has declined by a factor 2 in isotropic
luminosity, so we
expect an uncertainty of order this factor in our final estimate for
the upper limit to the intergalactic dust density. Averaging over
several observations (objects or epochs) would reduce this uncertainty.

In the absence of any information on the grain size
distribution, we will simply assume for now that the grains are
spherical, with a characteristic radius of order $a \sim 1 \mu$m.
We note that grains of this typical dimension remain optically thin to
X-ray absorption for photon energies $E \gax 1$ keV. 

Finally, we assumed a power law spectral shape for the X-ray emission 
of the quasar, using the slope we determined from a direct fit to the measured spectrum
(see Section 2), and we integrated the photons arriving at the telescope
over the telescope/ACIS-S effective area between 300 and 8000 eV, for direct comparison with the 
X-ray image. 
Since the halo is much
wider than the telescope response, we did not convolve it with the
telescope PSF. 

Our expressions for the scattering optical depth are based on the Rayleigh-Gans
approximation for the scattering cross section (Mauche \& Gorenstein 1986). At low
photon energies and large particle sizes, this approximation breaks down, and 
exact Mie theory should be used. For a characteristic particle size of 1 $\mu$m,
the Rayleigh-Gans approximation starts to fail below photon energies of order 1 keV. 
This implies that for scattering of low energy photons, the cross section
is overestimated by a factor up to a few (Smith \& Dwek 1998). In view of the 
approximations listed above, which introduce an uncertainty of the same order, we chose 
to perform the calculations based on the simple Rayleigh-Gans formulation. We also note
that due to the redshifting, the average photon energy at the scattering site is higher than
at the observer, which tends to dilute the error introduced by the use of the Rayleigh-Gans approximation.

We compare the model halo directly to the measured 
angular intensity profile of our source in Figure 2. Superimposed
on the observed profile of the quasar is a model halo
for an assumed dust density of $\Omega_d = 2 \times 10^{-6}$ (and 
$\Omega_m = 0.3$, $\Omega_{\Lambda} = 0.7$, grain radius $a = 1 \mu$m, density of the
grain material $\rho = 3$ g cm$^{-3}$).
For reference, on the scale of this figure, the background level in the image was approximately a constant 
0.04 counts pixel$^{-1}$ before subtraction. 
A dust halo at the chosen level would have
been detected significantly over the range
$\sim 10-100$ pixels: the net source count
in the image over the range $8-100$ pixels
is 107, while the halo model (excluding the inner 8 pixels)
contains 484 counts; the expectation value for the constant
background in this area is 2500 counts, with an expected 
Poissonian fluctuation of 50 counts.
We chose the dust density of $\Omega_{d} =
2 \times 10^{-6}$ somewhat arbitrarily. At this level, a halo is
clearly visible, while the dust density is already a factor of several below 
the astrophysically interesting fiducial level of $\Omega_d = 10^{-5}$ (see below). 

A more sensitive upper limit could be probably be obtained
by formally fitting model halos for various assumed sets of parameters,
taking the point source, background, and halo photon statistics
explicitly into account. However, at that point we should also properly
consider the effect on the upper limit on $\Omega_d$ of varying the 
physical parameters of the dust model (grain size and shape distribution, spatial
distribution, composition, etc.). Given the large uncertainties in these
parameters, we prefer to state the upper limit for a given fixed set of
dust parameters. Of these, the dependence on dust grain size is probably
the strongest.
To very rough approximation, the integrated halo intensity 
will scale inversely proportional with the dust grain radius. In addition, the shape of 
the halo changes with changing values for the radius (a
smaller grain radius will produce a wider halo at lower
amplitude, while a larger grain radius will produce the opposite
effect). As a result, the upper limit to $\Omega_d$ depends not only on the 
integrated halo intensity, but also on the background in the X-ray image. Nevertheless,
for grain radii not very different from 1 $\mu$m, the upper limit on $\Omega_d$ will
scale approximately inversely with grain radius. This is illustrated in Figure 2, where
we plot a model halo for assumed grain size $a = 0.25\ \mu$m, and $\Omega_d = 2 \times 10^{-6}$.

\section{Discussion}

Our upper limit of $\Omega_d < 2 \times 10^{-6}$
to the density of grey, smoothly distributed
intergalactic dust has several important astrophysical implications.
First, for the simplest assumptions for the properties of the
dust (spherical particles of characteristic radius 1 $\mu$m, of
constant comoving density), the upper limit
is about an order of magnitude below the dust density
required to explain the appearance of the $z < 1$ SN Ia Hubble
Diagram (Riess et al. 1998; Perlmutter et al. 1999) solely in terms of intergalactic extinction.
This information is now probably largely redundant,
since evidence has now become available to indicate that the apparent relative
dimming
of Supernovae (relative to a coasting Universe) 
for $z < 1$ turns into a relative brightening at large 
redshifts (Riess et al. 2004); and this of course 
very strongly suggests that dust extinction does not 
have a major effect. But apart from its 
providing independent confirmation of the relative unimportance of 
grey dust extinction on the overall appearance of the SN Ia
Hubble Diagram, the X-ray dust scattering halo technique can in principle
be refined to constrain and quantify systematic effects of 
dust extinction on attempts to use SN Ia to measure the equation of
state of Dark Energy.

Second, the limit already appears to rule out the most optimistic
estimates for the density of dust expelled by galaxies. Aguirre (1999)
estimates, on the basis of the measured cosmic metal production rate,
and the observed metallicity of galaxies and intracluster medium,
that as much as $\Omega_d \gax 2 \times 10^{-5}$ could
be smoothly distributed in the space between the galaxies; this is
not observed. Our result
may thus be used to put significant constraints on the cosmic dust
production rate and dispersal mechanism.

We note that our
finding appears to be at variance with the conclusion reached by
Windt (2002) that a dust halo would be much too faint to be
detectable. Comparing the integrated intensity quoted in that paper
for the
benchmark case ($\Omega_d = 4.5 \times 10^{-5}$,
$\Omega_m = 1$, $a = 0.1
\mu$m, $E = 1$ keV), we find that it is inconsistent with the
robust estimate
based on the optical depth (Eq. (2)), with the optical depth argument
giving a
fifteen times larger estimate. Also, Windt evaluated the
observability of a
halo for a particularly unfavorable choice of parameters ({\it e.g.},
effectively no dust beyond $z=0.5$, and constant physical, rather than
comoving, dust density).
Our estimate for the dust halo brightness based on a simple
optical depth
argument is robust, and we conclude that a halo should have been
detected around our redshift 4 quasar
if the dust density were significantly higher than
$\Omega_d \sim $ {\rm few}\ $\times 10^{-6}$. 

\acknowledgements

We gratefully acknowledge help from, and conversations with
Zoltan Haiman, Masao Sako, Ali Kinkhabwala, David Windt, Meg Urry, and 
Dani Maoz.
This work was supported by Chandra X-ray Observatory Grant
GO1-2144x from the Smithsonian Astrophysical Observatory.

\newpage

\noindent
Figure Captions:

\figcaption{
The ACIS/S3 image of QSO1508+5714 in the 0.3$-$8 keV band, binned in 0.5
arcsec pixels. North is up, East is to the left. The color scaling
is logarithmic, with the brightest pixel containing 1370 photons,
and the dimmest non-black pixels containing one photon each. The
position of the brightest pixel is at RA (J2000) = $15^{\rm h}
10^{\rm m} 02^{\rm s}.90$, DEC (J2000) = $+57^{\rm d} 02\arcmin
43\arcsec.3$. The image is clearly extended towards the southwest.
}

\figcaption{
The azimuthally averaged, background subtracted $0.3-8$ keV
intensity profile of the quasar, in
counts/surface area (area measured in pixels 0.5 arcsec on a side), vs. radial
coordinate ({\it points with error bars}).
Superimposed are a ray trace simulation for the core of the image
({\it dotted line}), and a radial profile for the bright source 3C273
({\it solid jagged line}); the downturn at $r < 5$ pixels (2.5
$\arcsec$) is a consequence of photon pileup in the detector. The upper
solid curve represents a simple dust scattering halo model
corresponding to an average dust density of $\Omega_d = 2 \times
10^{-6}$, for dust particle size 1 $\mu$m. The flat distribution extending out to
approximately 300 pixels is a model halo for particle size $0.25\ \mu$m and 
$\Omega_d = 2 \times 10^{-6}$.
}

\vfill\eject

\begin{figure}
\epsscale{1}
   \plotone{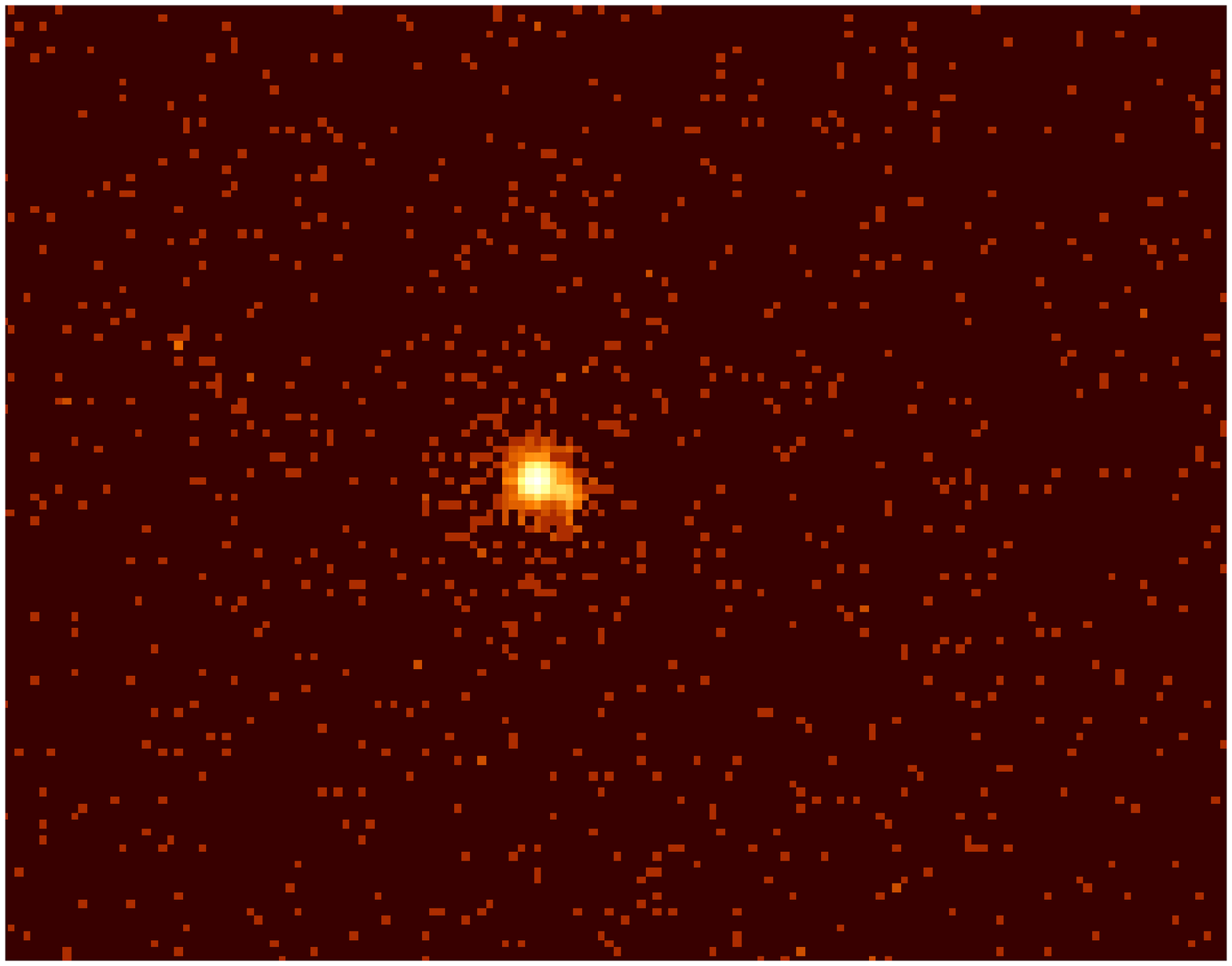}
\end{figure}

\vfill\eject

\begin{figure}
\epsscale{1}
   \plotone{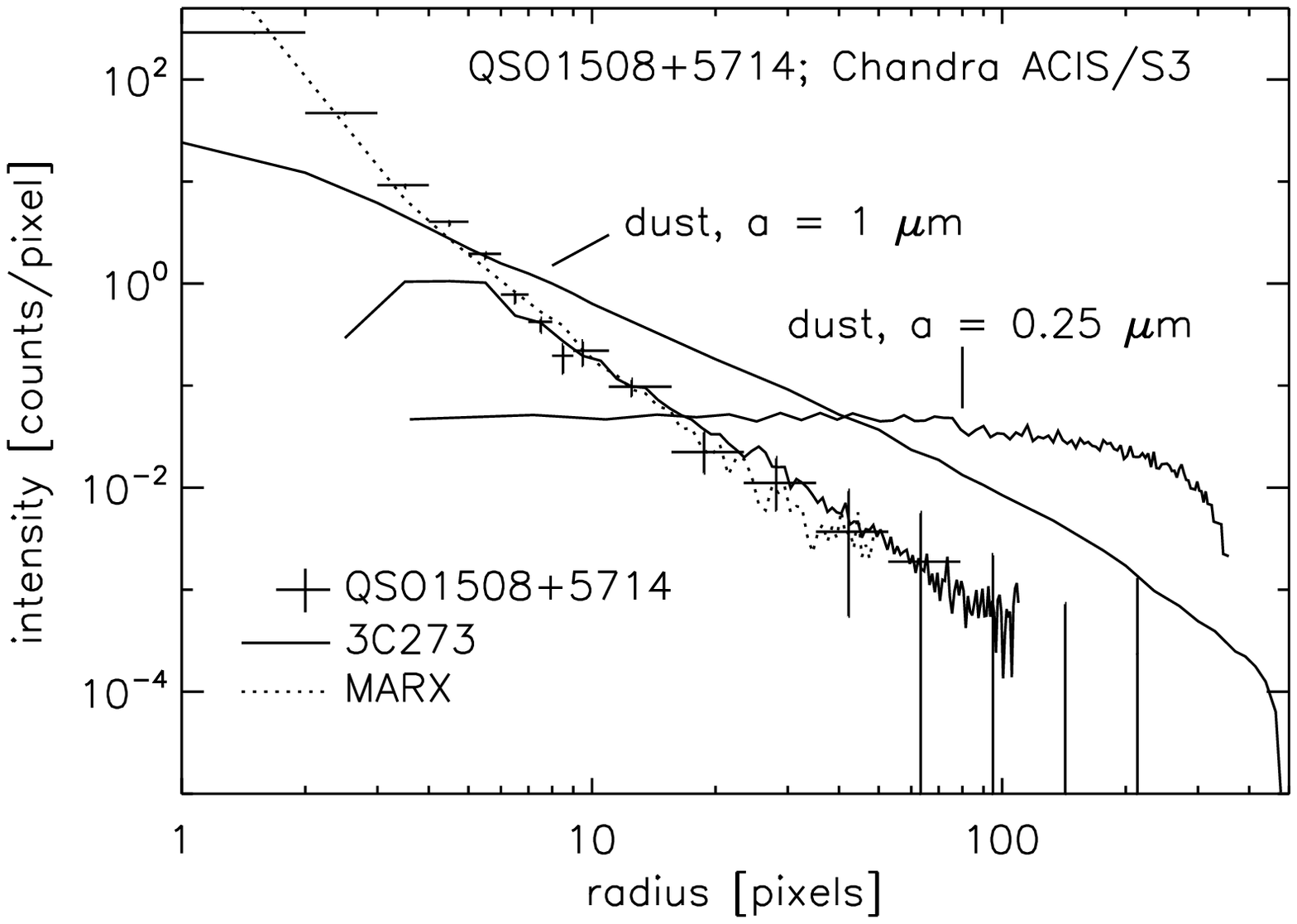}
\end{figure}

\end{document}